\documentclass[a4paper,reprint,3p,times,onecolumn,12pt]{elsarticle} 
\usepackage{amsmath,amsthm}
\usepackage{amsfonts}
\usepackage{amssymb}
\usepackage{hyperref}
\usepackage{graphicx}
\usepackage{latexsym}
\usepackage{amssymb}
\usepackage{amscd,amssymb}
\usepackage[arrow,matrix]{xy}
\usepackage{graphicx}
\usepackage{epstopdf}

\usepackage{soul}
\biboptions{sort&compress}
\usepackage{setspace}
\usepackage{amscd,amsmath,amsthm,amssymb,color}
\usepackage{xcolor}
\def  \cb{\color{black}}
\allowdisplaybreaks
\theoremstyle{plain}

\emergencystretch=\maxdimen
\newcommand{\bea}{\begin{eqnarray}}
\newcommand{\eea}{\end{eqnarray}}
\newcommand{\bes}{\begin{subequations}}
	\newcommand{\ees}{\end{subequations}}

\usepackage{fancyhdr}
\let\today\relax
\makeatletter
\def\ps@pprintTitle{%
	\let\@oddhead\@empty
	\let\@evenhead\@empty
	\def\@oddfoot{Journal Ref.: Chaos Solitons \& Fractals {\bf 167} (2023) 113058. \hfill DOI: \href{https://doi.org/10.1016/j.chaos.2022.113058}{10.1016/j.chaos.2022.113058}}%
	\let\@evenfoot\@oddfoot
}
\makeatother

\begin{document}
	\title{{Lump and Soliton on Certain} Spatially-varying Backgrounds\\ for an Integrable {(3+1) Dimensional} Fifth-order\\ Nonlinear Oceanic Wave Model} 
	
	\author[nit]{Sudhir Singh}
	\author[apctp]{K. Sakkaravarthi \corref{cor}}
	\author[nit]{K. Murugesan}
	
	\address[nit]{Department of Mathematics, National Institute of Technology, Tiruchirappalli -- 620015, Tamil Nadu, India}
	\address[apctp]{Young Scientist Training Program, Asia-Pacific Center for Theoretical Physics (APCTP),\\ POSTECH Campus, Pohang -- 37673, South (Republic of) Korea}
	
	\cortext[cor]{{\noindent Corresponding Author: ksakkaravarthi@gmail.com; karuppaiya.sakkaravarthi@apctp.org (K. Sakkaravarthi)}\newline 
		 \indent \quad  Email: sudhirew@gmail.com (Sudhir Singh)\newline   
		 \indent  \quad  Email: murugu@nitt.edu (K. Murugesan)}
	
	\journal{Chaos, Solitons \& Fractals}
	\date{\today}
	\setstretch{1.20}	
	\begin{abstract} 
		The dynamics of nonlinear waves with controllable dispersion, nonlinearities, and background continues to be an exciting line of active research in recent years. In this work, we focus to investigate an integrable (3+1)-dimensional nonlinear model describing the evolution of {water waves} with higher-order temporal dispersion by characterizing the dynamics of lump and soliton waves on different spatially-varying backgrounds. Particularly, we construct some explicit lump wave and kink-soliton solutions for the considered model through its trilinear and bilinear equations with appropriate forms of the polynomial and exponential type initial seed solutions, respectively. Additionally, we obtain hyperbolic and periodic waves through bilinear B\"acklund transformation. We explore their propagation and transformation/modulation dynamics due to different spatial backgrounds through categorical analysis and clear graphical demonstrations for a complete understanding of the resulting solutions. Our analysis shows that the lump solution results into a simultaneous existence of well-localized spike and declining (coupled bright-dark wave) structures on a constant background, while the soliton solution exhibits kink and anti-kink wave patterns along different spatio-temporal domains with controllable properties. When the arbitrary spatial background is incorporated appropriately, we can witness the manifestation of soliton into manipulated periodic and interacting dynamical structures. On the other hand, the lump wave results into the coexistence, interacting wave and breather formation due to periodic and localized type arbitrary spatial backgrounds. The present results will be an important addition to the context of engineering nonlinear waves with non-vanishing controllable backgrounds in higher-dimensional models. Further, the present study can be extended to investigate several other nonlinear systems to understand the physical insights of the variable background in their dynamics.
		\\ 
		
		\begin{keyword}{ Oceanic waves; (3+1)D nonlinear evolution equation; Lump wave; Kink soliton; Jacobi elliptic function; Spatially-varying background.} 
		\end{keyword}
	\end{abstract}
	
	\maketitle
	\setstretch{1.10}	
	\newpage
	{\bf Highlights}
	\begin{itemize}
		\item Investigated an integrable (3+1)-dimensional nonlinear oceanic wave model.
		\item Constructed explicit lump and soliton solution through trilinear and bilinear formalisms.
		\item Analysed the dynamics of nonlinear waves on different spatially-varying backgrounds.
		\item Explored the influence of backgrounds in the transformation of soliton and lump structures.
		\item Observed the coexistence, superposition, breather formation and interaction of different waves.
	\end{itemize}
	
	\setstretch{1.20}	
	\section{Introduction}
	Investigations on various nonlinear wave structures that include
	solitons/solitary waves, kinks, breathers, shocks, rogue waves, lumps,
	dromions, peakons, cuspons, compactons, periodic and elliptic waves, etc.
	emerge an interesting avenue among the researchers in the past few
	decades~\cite{Yang-book}. This is accomplished through the development of
	different analytical methodologies to solve the underlying evolution equations
	and numerical simulations running on efficient computation~\cite{ML-book}. 
	Solitons are one of the inevitable and remarkably stable nonlinear waves that are being a front-runner with multidisciplinary applications starting from 
	the data transmission networks and nonlinear optics to Bose--Einstein
	condensate, plasma systems, and so on~\cite{bgo,bgu,boris,ksa}. The thorough
	comprehension of many real-world problems in plasmas, fibre optics, fourth and
	fifth states of matter, etc. is significantly advanced by the mathematical and
	physical analysis of localized wave solutions and long-time asymptotic
	behaviours of these soliton models~\cite{ML-book,Yang-book,bgo,bgu,boris,ksa}.
	Further, on contrary to the solitons, rogue waves are another type of nonlinear
	entities that are first observed as deep oceanic waves with huge instability
	and they are observed in various fields~\cite{akm-rogue}. With certain structural
	similarities and connection with unstable rogue waves, stable propagating
	dromions and lumps are observed in several higher-dimensional models arising in
	the context of metamaterials~\cite{zba}, flexural-gravity waves~\cite{mra},
	ferromagnetic film~\cite{xwj}, water waves~\cite{wx1}, superfluids~\cite{djf}
	and in hydrodynamics~\cite{fba}.
	
	In the context of water wave theory, the complete study on hydrodynamical and
	related physical systems are achieved by exploring several integrable as well
	as nonintegrable evolution equations in one and higher
	dimensions~\cite{Yang-book}. Integrability is a fascinating property to
	characterize any dynamical models in addition to existence of Lax pair and
	infinitely many conserved quantities~\cite{ML-book}. The higher dimensional  
	generalization of the well-known one-dimensional soliton models gives a more
	clear picture of the evolutionary behaviours of the associated nonlinear
	waves~\cite{hdg,gqx,ylm,lka}. 
	Integrable soliton models consisting of first-order temporal evolution $u_{t}$ are observed to describe unidirectional wave propagation, that includes Korteweg--de Vries (KdV) and modified-KdV type equations. Further, nonlinear models such as Boussinesq equation, Klein--Gordon equation and Kaup--Kupershmidt equation containing second-order temporal evolution $u_{tt}$ are the manifestation of bidirectional propagation of waves. 
	Interestingly, nonlinear models with the third-order temporal evolution
	$u_{ttt}$ are useful extension~\cite{amw1,amw2,gwa,amw3,gqx1,rip1} in a
	variety of physical circumstances where the respective causalities are
	preserved~\cite{jgr,csc}. The integrability properties and nonlinear wave
	structures in this type of nonlinear models are comparatively less explored. 
	
	Recently, Wazwaz introduced the following (3 $+$ 1)-dimensional fifth-order
	nonlinear model with third-order temporal and spatiotemporal dispersion
	effects~\cite{amw4}:
	\begin{equation}
	u_{ttt}-u_{txxxx}-12 u_{xt} u_{xx}-8 u_{x}u_{xxt}-4 u_{t}u_{xxx}+(\alpha u_{x}+\beta u_{y}+\gamma u_{z})_{xx}=0. \label{t1}
	\end{equation}
	Further, the above model is found to be integrable  through Painlev\'e
	analysis and soliton solutions are obtained using a simplified Hirota
	method~\cite{amw4} along with different wave solutions of its one-dimensional
	counterpart models~\cite{gqx1,rip1}. 
	
	Though solving these nonlinear equations is analytically a challenging task in
	general, there are significant developments in formulating powerful tools such
	as inverse scattering transforms, Hirota's bilinear method, Darboux and
	B\"acklund transforms, Lax pairs, Riemann--Hilbert approach, KP hierarchy
	reduction, Gauge transformation, Lie symmetry method,
	etc.~\cite{Yang-book,ML-book,bgo,bgu}. 
	Apart from these, {similarity reduction approach~\cite{cp1,cp2,cp3,cp4}} and
	various ancillary methods involving direct algebraic techniques (ODE reduction
	approaches), auxiliary equation method, Kudryashov expansion method,
	Riccati--Bernoulli sub-ODE method, sinh-Gordon expansion method, cosh-tanh
	method, simplest equation method, etc. are utilized widely to get different
	classes of travelling wave solutions~\cite{amwb}. 
	Especially, these methodologies provide a variety of exotic wave patterns,
	including solitons, breathers, lumps, dromions, rogue waves, and elliptic
	waves, and limitations are arising in utilizing strong tools like inverse
	scattering transform, Riemann--Hilbert approach and Darboux transformation due
	to the lack of Lax pair guaranteeing complete integrability for many of the
	higher-dimensional soliton models. On the advantageous part, the Hirota
	bilinear method is an intermediate tool which can be utilized to extract the
	localized nonlinear wave solutions to most of the integrable as well as a few non-integrable soliton models and it becomes a widely used tool to
	obtain several localized nonlinear wave solutions~\cite{Hirota-book}. {
		Through $N$-soliton solutions, utilizing the long-wave limit
		method, one can obtain different localized waves including lumps, rogue waves,
		breathers and different types of interacting waves. Recently, an improved
		long-wave limit method is proposed for obtaining higher-order rogue waves for
		the nonlinear Schr\"odinger equation~\cite{abc4}, which can also be extended to
		many other integrable models.}
	{ The Hirota bilinear method  is directly linked with truncated Painlev\'e
		approach, where the later is successfully utilized to a variety of constant and
		variable coefficient soliton models to identify their integrability property and to obtain various nonlinear wave
		structures through auto-B\"acklund and hetero-B\"acklund transformations
		including Boussinesq--Burgers system~\cite{cl1}, Whitham--Broer--Kaup-like
		system~\cite{cl2}, variable coefficient generalized Burgers system~\cite{cl3}
		and the generalized variable-coefficient KdV-modified KdV equation~\cite{cl4}.} 
	{  Particularly, this Hirota method is successfully used to identify certain
		interesting dynamics of multi-lumps and lump chains in the framework of the
		KP$1$ model~\cite{abc1,abc2,abc3}}.
	{  Interestingly, in addition to the above approaches, some recent
		methodologies such as probabilistic approaches~\cite{ch1,ch2} and deep learning
		techniques~\cite{ch3,ch4,ch5} attract much interest in understanding different
		nonlinear systems in different perspectives.}
	
	{Localized nonlinear waves on variable backgrounds are another important aspect
		of study in recent times across multiple disciplines starting from optics
		to hydrodynamics systems.   However, we can find only a handful amount of works
		for such variable background nonlinear waves compared to their counterparts on
		constant backgrounds. In the context of nonlinear optics, the effects of
		spatially- and temporally-varying backgrounds on solitons, breathers and rogue
		waves have revealed interesting outcomes that can be useful to understand the
		influence of nonuniform models~\cite{ksjpa20,ksps20,ksa}. Similarly, the
		dynamics of nonlinear waves on controllable backgrounds due to non-autonomous
		nonlinearities are explored in Bose--Einstein condensates
		too~\cite{boris,tk1,tk2,tkan,mani1,mani2}. { Recently, some studies are
			reported for nonlinear wave structures in certain higher-dimensional nonlinear
			models describing shallow or deep water waves, which include the analyses of
			breathers in a variable-coefficient (3$+$1)D shallow water wave
			model~\cite{pfh,epjp19}, rogue waves and solitons on spatio-temporally variable
			backgrounds in (3$+$1)D Kadomtsev--Petviashvili--Boussinesq system~\cite{ssi},
			solitons in an extended (3$+$1)D shallow water wave equation~\cite{jgl}, and
			interaction waves in both (3$+$1)D and (4$+$1)D Boiti--Leon--Manna--Pempinelli
			models~\cite{pfha,epjp21} to name a few.} 
		
		The above studies revealed different exciting results such as amplification,
		compression, bending/snaking, tunnelling/cross-over, superposed structures,
		etc.~\cite{ksjpa20,ksps20,ksa,boris,tkan,tk1,tk2,mani1,mani2,pfh,epjp19,epjp21,jgl,ssi,pfha}.
		So, it is our natural interest to question the possibility of realizing such
		phenomena in the above mentioned fifth-order nonlinear model~\eqref{t1}. To the best of
		our knowledge, the Painlev\'e integrable model \eqref{t1} under consideration
		is only solved for solitary waves on a constant background. 
		Motivated by the above interesting observations, with a special emphasis on higher-dimensional integrable soliton models, here in this work, our objective is to study the effects of controllable spatial backgrounds on the localized nonlinear waves (lumps and solitons) of the (3 $+$ 1)D fifth-order integrable soliton model with third-order temporal dispersion as given by Eq.~\eqref{t1}. Particularly, we are aimed to demonstrate the dynamics by adopting both exponential and rational localized waves on spatially-varying backgrounds.}

	In view of the above perspective, the remaining part of the manuscript is arranged in the following manner. We construct the lump solution of the considered model \eqref{t1} in Sec. \ref{sec2} through its trilinear form and polynomial type seed solution along with a detailed study of its dynamics on constant as well as spatially-varying backgrounds. Section \ref{sec3} deals with the soliton solution of Eq.~\eqref{t1} using Hirota's bilinearization method and their detailed evolutionary dynamics is investigated. In Sec. \ref{sec5}, the bilinear B\"acklund transformation is derived. The final  \ref{sec-conclusion} is allotted for conclusions of the present work along with certain future outlook.  
	
	\section{Trilinear Equation and Lump Wave Dynamics on Background}\label{sec2}
	{ In this section, we aim to analyse the dynamics lump wave arising in the above-considered model (\ref{t1}) on the constant non-vanishing background as well as spatially-varying controllable backgrounds through its exact solution.} For this purpose, first, we implement the following superposed bilinear logarithmic transformation:
	\begin{equation}
	u(x,y,z,t)=u_{0} (y,z)+\left[{\ln}(f(x,y,z,t)\right]_{x}, \label{t2}  
	\end{equation}
	where $u_{0} (y,z)$ is an arbitrary background which depends on two spatial dimensions and $f (x,y,z,t)$ is the required function to be calculated for constructing the solution of Eq. (\ref{t1}). 
	{\cb Here we wish to emphasize that a simple form of superposed bilinear transformation with a one-dimensional space-varying $\psi(z)$ background is used to study breathers in a variable-coefficient (3+1)D shallow water wave model in \cite{pfh,epjp19}, while in Ref. \cite{pfha,epjp21} multiple interaction solutions for the (3+1)D and (4+1)D Boiti--Leon--Manna--Pempinelli equations with time-varying background $\psi(t)$ is presented. Recently, in Ref.~\cite{ssi}, we have generalized the approach and investigated the effects
		of arbitrary spatio-temporal background $\psi(z,t)$ background in the rogue waves and solitons for a (3+1)D Kadomtsev-Petviashvili-Boussinesq (KPB) equation. Next to that, there is another recent work \cite{jgl} on kink-solitons for an extended (3+1)D shallow water wave equation with a similar $\psi(z,t)$ background function. Following these exciting outcomes, in the present work, our aim is to study the influence of pure arbitrary two-dimensional spatial background in the evolution
		of lump and soliton, which shows different phenomena as
		explained in the forthcoming part of the manuscript. As a future study, one can attempt to apply this and other generalized arbitrary backgrounds (like $\psi(x,y,z,t)$ as hinted in our previous work \cite{ssi}) to different nonlinear wave equations supporting a variety of wave structures that can reveal excellent features and applications.}
	
	Upon substituting the above superposed transformation (\ref{t2}) into the considered model Eq. (\ref{t1}), the following trilinear form is obtained as as follows.
	\bea
	&T(f)\Rightarrow& 2f_{t}^3+2 f_{x}[2 f_{xx} f_{xt}+f_{x}(\alpha f_{x}+\beta f_{y}+\gamma f_{z}-2 f_{xxt})]-f[f_{xx}(\beta f_{y}+\gamma f_{z}+2 f_{xxt}) \nonumber\\ && +f_{x}(3\alpha f_{xx}+2\beta f_{xy}  +2\gamma f_{xz}-4 f_{xxxt})]	+f_{t}[2 (f_{xx}^2-2 f_{x}f_{xxx})+f(f_{xxxx}-3 f_{tt})] \nonumber\\ &&+f^2(\alpha f_{xxx}+\beta f_{xxy}+\gamma f_{xxz}+f_{ttt}-f_{xxxxt})=0. \label{t3}
	\eea 
	{\cb Noted that the above trilinear form (\ref{t3}) is surprisingly independent of $u_0(y, z)$ under the superposed transformation (\ref{t2}). This ensures that $u_0(y,z)$ is arbitrary and different types of backgrounds including elliptic functions can be introduced to any possible nonlinear wave structures to the considered fifth-order integrable soliton model (\ref{t1}), once such solutions $f(x,y,z,t)$ satisfy (\ref{t3}). Considering the length of the article, in this section, we demonstrate the algorithm with only the first-order lump solution.}  
	
\begin{figure}[h]
		\centering\includegraphics[width=0.99\linewidth]{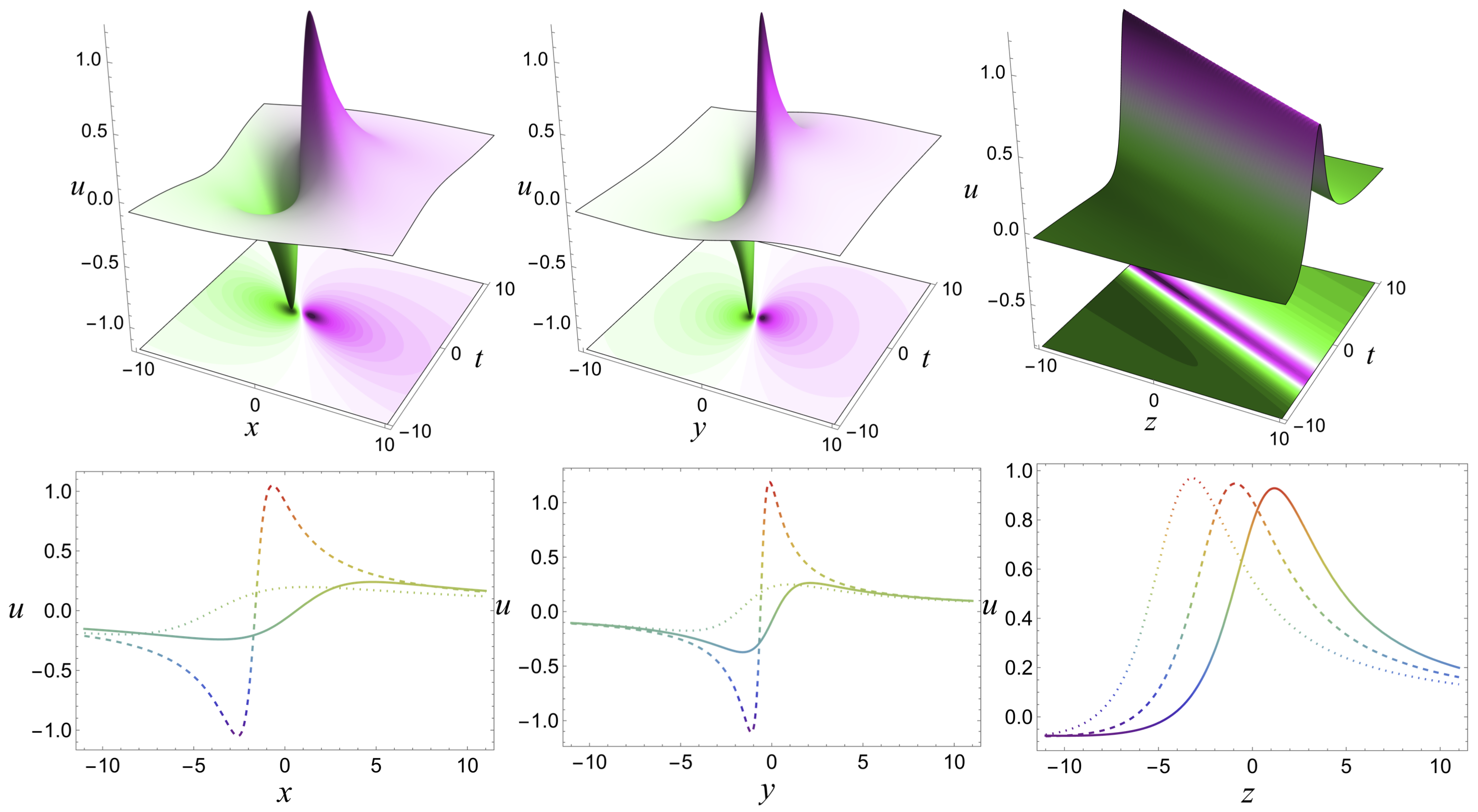}
		\caption{The nature of lump soliton along different spatio-temporal planes without any background $u_0=0$ are given in the 3D plots for the choice $\delta_1=1.0$, $\delta_2=0.5$, $\theta_0=0.5$, $\theta_1=0.3$, $\theta_2=0.7$, $\gamma_1=0.45$, $\gamma_2=0.25$, $\alpha=0.4$, $\beta=0.9$, and $\gamma=1.3$. Especially, doubly-localized lump structure along $x-t$ at $y=z=0.4$ and $y-t$ at $x=z=0.4$, while we obtain line soliton along $z-t$ at $x=y=0.4$. The lower 2D plots represent their corresponding structure at $t=-1.5$ (solid line), $t=0$ (dashed line), and $t=1.5$ (dotted line).}
		\label{fig-lump-1}
	\end{figure}

\begin{figure}[h]
		\centering\includegraphics[width=0.99\linewidth]{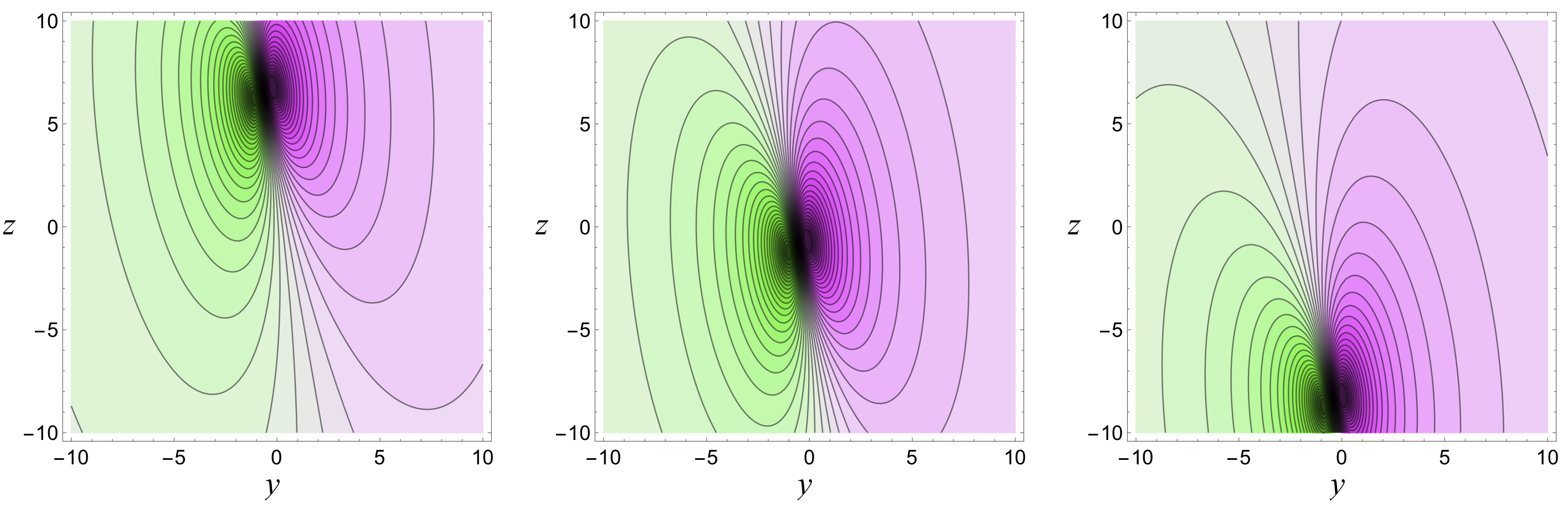}
		\caption{The contour plot of the propagation (localization movement) of the lump structure along the two spatial plane $y-z$ at different times $t=-3.4,0,3.4$ and $x=0.4$ without any background for the same choice as in Fig. \ref{fig-lump-1}.}
		\label{fig-lump-2}
	\end{figure}

	To construct lump solution, we choose the following quadratic function as seed solution \cite{wx1}
	\begin{equation} 
	f(x,y,z,t)=\theta_{0}+(\alpha_1 x+\beta_1 y+\gamma_1 z+\delta_1 t+\theta_1)^2+(\alpha_2 x+\beta_2 y+\gamma_2 z+\delta_2 t+\theta_2)^2, \label{t4}
	\end{equation} 
	where $\theta_i, \alpha_j, \beta_k, \gamma_l, \delta_m, i=0,1,2, j=k=l=m=1,2$, are the parameters of the lump solution. Substituting (\ref{t4}) in (\ref{t3}), we get a polynomials in $x,y,z$ and $t$. Further, equating to zero all linearly independent terms to zero, after solving the system, we get the following constraints over lump parameters as 
	\begin{equation}
	\alpha_1=0, \alpha_2=\sqrt{\delta_1} \theta_{0}^{1/4}, \beta_1=\dfrac{\delta_1^2-3\delta_2^2-\gamma \gamma_1 \sqrt{\theta_0}}{\beta \theta_0}, \beta_2=\dfrac{\delta_2^3-3 \delta_1^2\delta_2+\gamma \gamma_2 \delta_1 \sqrt{\theta_0}+\alpha \delta_{1}^{3/2}\theta_{0}^{3/4}}{-\beta \delta_1 \sqrt{\theta_0}}.\label{t5}
	\end{equation}
	Hence, we shall obtain the required form of $f$ from Eq. (\ref{t4}) by substituting the quantities obtained in Eq. (\ref{t5}), from which we can deduce the resultant lump wave solution $u$ from Eq. (\ref{t2}) in a compact form as follows.
	\begin{equation}
	u(x,y,z,t)=u_0 (y, z)+\frac{2\sqrt{\delta_1} \theta_{0}^{1/4}(\sqrt{\delta_1} \theta_{0}^{1/4} x+\beta_2 y+\gamma_2 z+\delta_2 t+\theta_2)}{\theta_{0}+(\beta_1 y+\gamma_1 z+\delta_1 t+\theta_1)^2+(\sqrt{\delta_1} \theta_{0}^{1/4} x+\beta_2 y+\gamma_2 z+\delta_2 t+\theta_2)^2}. \label{lump-sol}
	\end{equation}
	The other parameters are given in the Eq. (\ref{t5}). The above obtained lump solution is characterized by seven arbitrary solution parameters ($\delta_1$, $\delta_2$, $\theta_0$, $\theta_1$, $\theta_2$, $\gamma_1$, $\gamma_2$)  and three arbitrary system parameters ($\alpha$, $\beta$, and $\gamma$) in addition to the arbitrary spatial background $u_0 (y, z)$. By tuning these arbitrary parameters/functions one can control the nature and evolution of the resulting lump solution. The present solution (\ref{lump-sol}) describes a doubly-localized lump with one peak and one well/hole on a constant background. The nature of the resulting lump wave admits different width, localization and position along different spatio-temporal domains, whereas it exhibits same amplitude throughout the evolution when there exits no background function ($u_0=0$). Note that the free parameters available in the solution (\ref{lump-sol}) play crucial role in controlling the identities of the lump such as the amplitude, width, localization and position, so that a rich categories of lump wave profiles can be constructed accordingly. The amplitude of the lump wave above the zero or non-zero background is proportional to the factor $2\sqrt{\delta_1} \theta_{0}^{1/4}$, while the other parameters help to manipulate other identities of the lump wave. For a clear understanding on the above discussion, we have given graphical demonstration of lump solution (\ref{lump-sol}) along different spatio-temporal domains in Fig. \ref{fig-lump-1}, while its dynamics along spatial dimension $y-z$ at three different time is shown in Fig. \ref{fig-lump-2}. It is important to highlight that the nature of lump wave remains unaltered and it takes different positions due to temporal evolution. 
	
	\begin{figure}[h]
		\centering\includegraphics[width=0.99\linewidth]{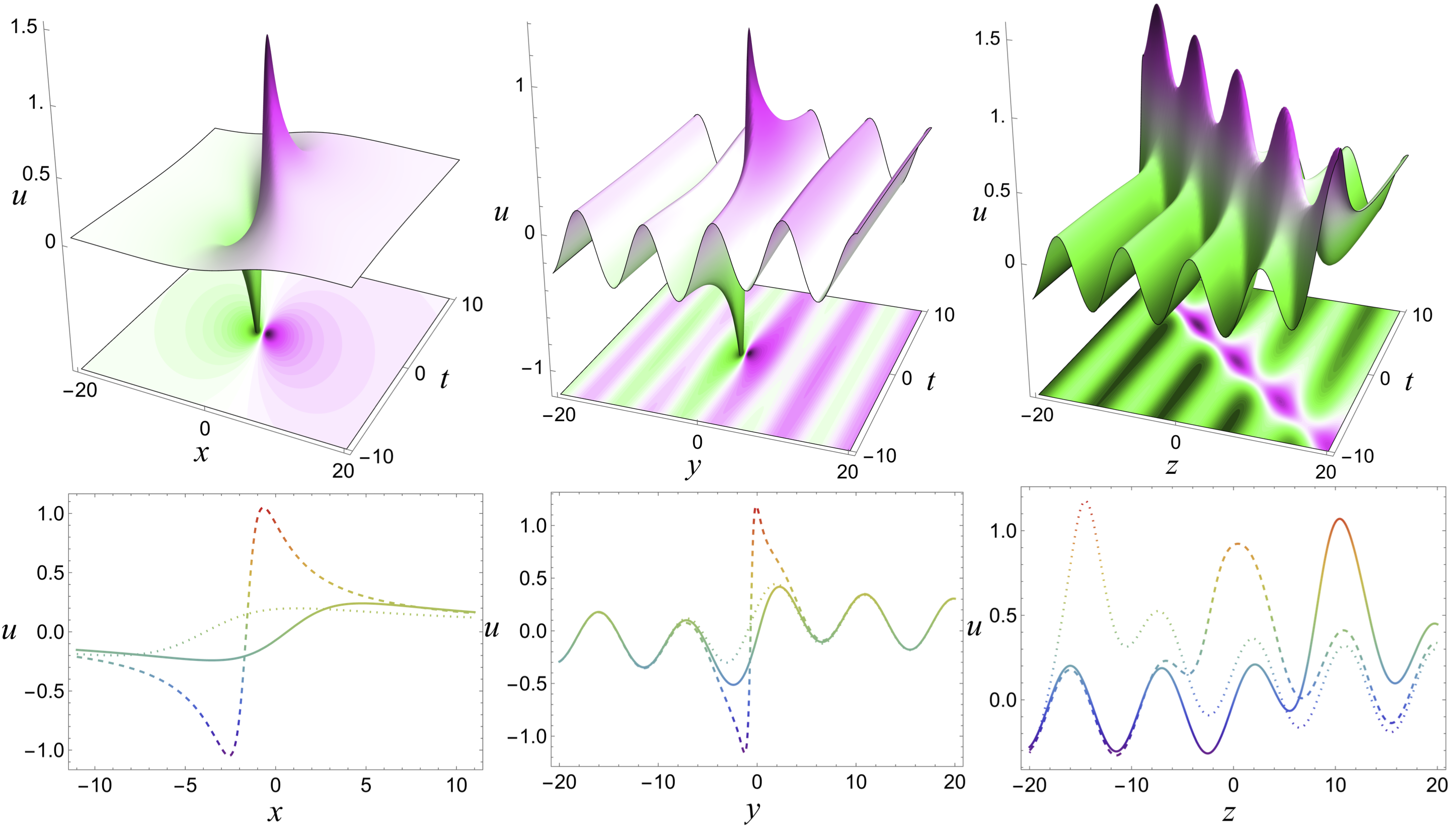}
		\caption{The impact of periodic background in the evolution of lump wave along different spatio-temporal planes for $u_0(y,z)=0.35~ \mbox{sn}(0.5 y+0.5 z,0)$ with other parameters as given in Fig. \ref{fig-lump-1}.} 
		\label{fig-lump-3}
	\end{figure} \begin{figure}[h]
		\centering\includegraphics[width=0.99\linewidth]{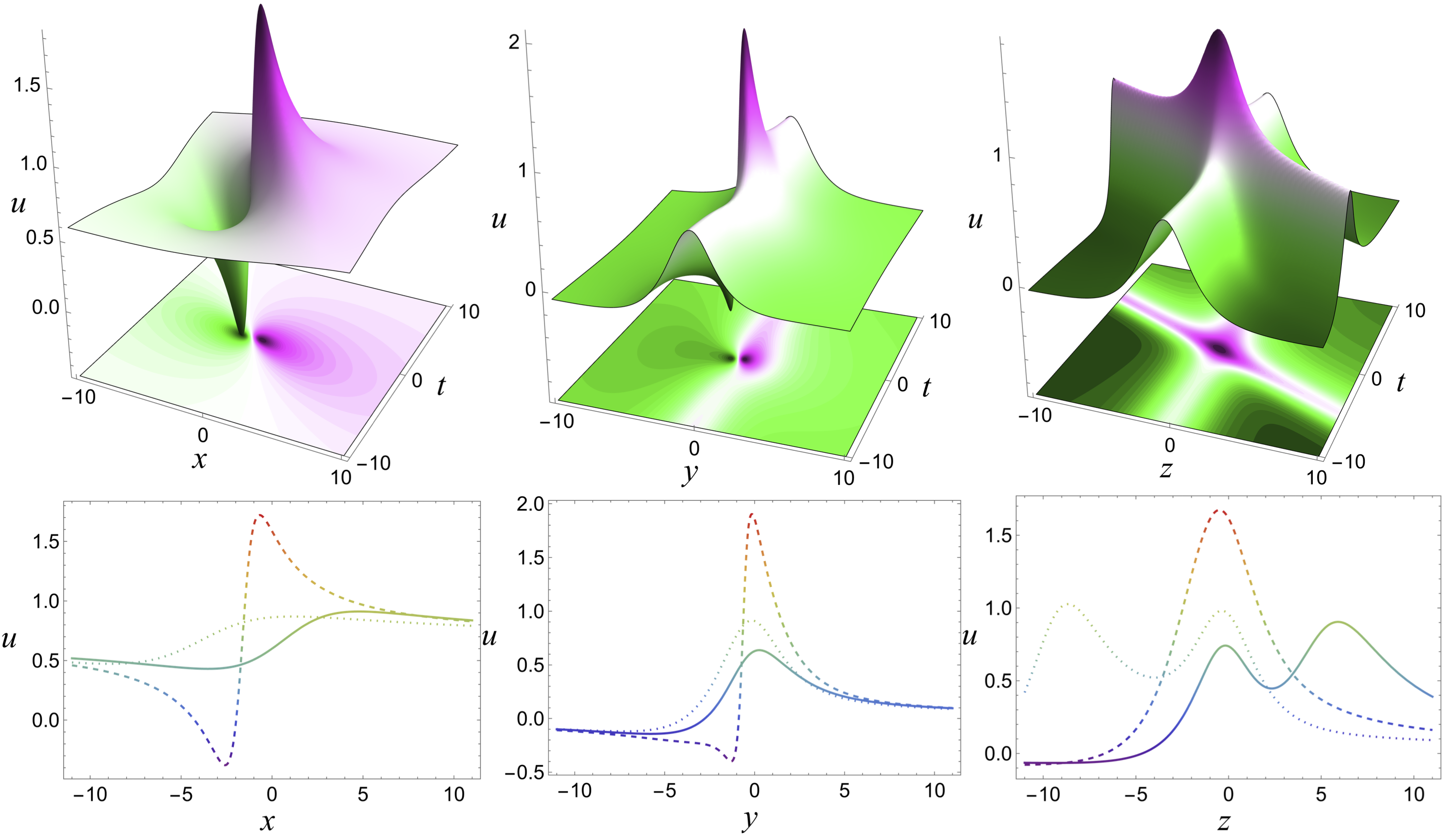}
		\caption{The impact of localized bell-type soliton background in the evolution of lump wave along different spatio-temporal planes for the same choice of parameters as given in Fig. \ref{fig-lump-1} with $u_0(y,z)=0.75~ \mbox{cn}(0.5 y+0.5 z,1)$.} 
		\label{fig-lump-4}
	\end{figure}
	\begin{figure}[h]
		\centering\includegraphics[width=0.99\linewidth]{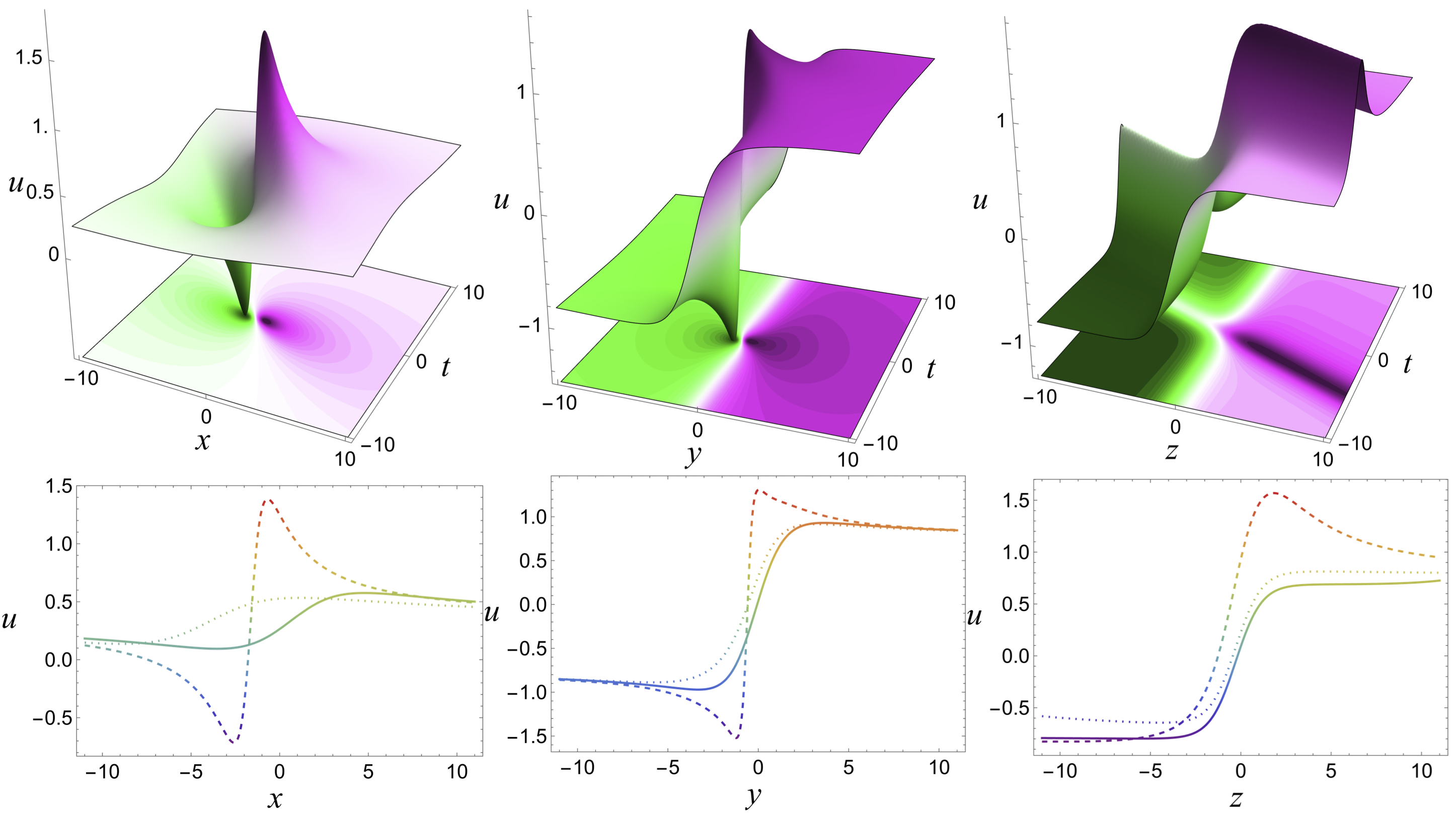}
		\caption{The impact of kink-type soliton background in the evolution of lump wave along different spatio-temporal planes for the same choice of parameters as given in Fig. \ref{fig-lump-1} with $u_0(y,z)=0.75~ \mbox{sn}(0.5 y+0.5 z,1)$.} 
		\label{fig-lump-5}
	\end{figure}
	\begin{figure}
		\centering\includegraphics[width=0.99\linewidth]{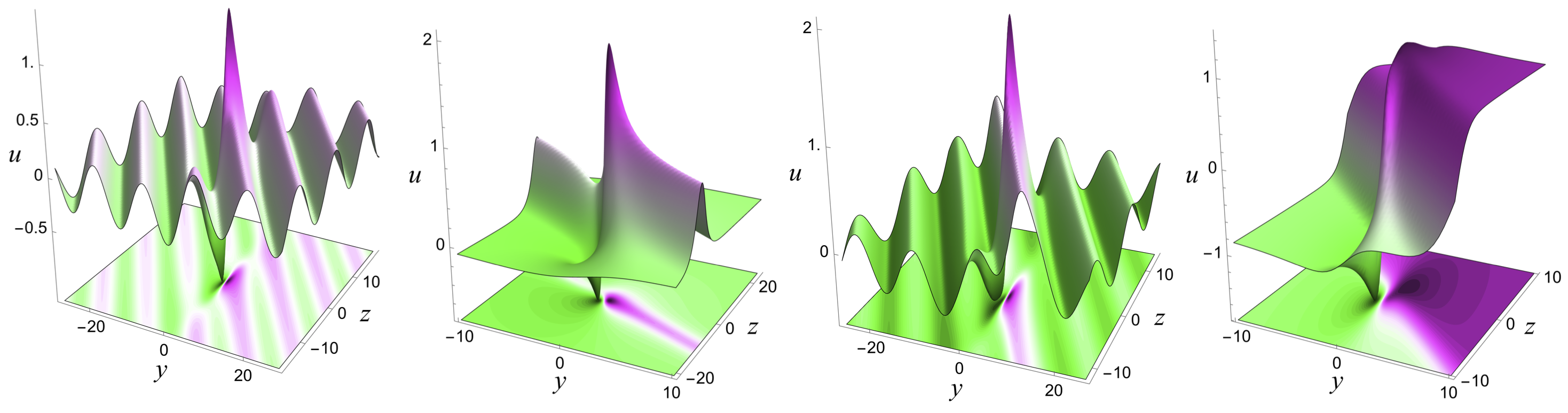}
		\caption{The manipulation of lump wave along the spatial domain $y-z$ due to periodic, localized bell-type soliton, combined bell-soliton with periodic, and kink-soliton backgrounds with parameter values as given in Figs. \ref{fig-lump-1}-\ref{fig-lump-5}.} 
		\label{fig-lump-6}
	\end{figure}

	Interestingly, when the arbitrary background $u_0(y,z)$ is taken into effect and non-vanishing, the lump wave undergoes various manifestation based on the nature of background variation and results into different wave phenomena. To shed light on the understanding, we have adapted two type of arbitrary spatial background function such as (i) localized soliton type background and (ii) periodic background by incorporating Jacobi elliptic functions as given below.
	\bea u_0(y,z)=p_1 \mbox{sn}(a_1 y+b_1 z,q_1)+p_2 \mbox{cn}(a_2 y+b_2 z,q_2),\label{jacobi} \eea
	where $p_j$, $a_j$, and $b_j$ ($j=1,2$) are arbitrary real constants, while $q_1$ and $q_2$ are the elliptic modulus parameters ($0\leq q_1,q_2 \leq 1)$, using which we shall incorporate periodic ($q_1=0$ or $q_2=0$), kink-like soliton ($q_1=1$ and $p_2=0$ ) and bell-type soliton ($q_2=1$ and and $p_1=0$) type backgrounds to the initial lump solution of Eq. (\ref{t1}). The function $u_0(y,z)=p_1 \mbox{sn}(a_1 y+b_1 z,0)$ manifests the lump to appear on the periodic background along $y-t$, whereas one can obtain an interaction between periodic wave and breather along $z-t$ plane. Additionally, the localized soliton type spatial background leads to the superposition lump on soliton along $y-t$, while it results into interacting solitons along $z-t$. However, both type of backgrounds show no impact on the lump along $x-t$ due to the fact that the spatial background $u_0(y,z)$ is considered to be on two dimensions only. For getting further insights to the above arguments, we have graphically shown the influence of both periodic and soliton backgrounds in Figs. \ref{fig-lump-3}-\ref{fig-lump-6}. Especially, we have given the influence of spatially-varying periodic background in Fig. \ref{fig-lump-3} along different spatio-temporal domain, which shows that the lump wave remains unaltered along $x-t$, while it is affected by the background along $y-t$ and transform into interaction with a periodic wave. Interestingly, one can observe the formation of breather along $z-t$ and its interaction with a background periodic wave. In the case of localized bell-type soliton background, we witness the superposition of the lump with solitary wave along $y-t$ and results into an passing-through collision between two solitary waves along $z-t$ as shown in Fig. \ref{fig-lump-4}. Figure \ref{fig-lump-5} depicts the impact of kink-type soliton background on the lump wave along different spatio-temporal planes. Finally, we have shown how the arbitrary controllable backgrounds modulate the lump wave along the spatial domain $y-z$ in Figs. \ref{fig-lump-6}.

	\section{Bilinear Formalism and Soliton Dynamics on Background} \label{sec3}
	In this section, we construct  the solitary wave solution by adopting bilinearization algorithm using Hirota derivatives \cite{Hirota-book} and study their dynamics under different backgrounds of interest. For this purpose, first, by applying the following dimension reducing transformation:
	\bea u(x,y,z,t) =  u(\eta , y, z), \eea 
	where $\eta = (x+\delta t)$, to the original Eq. (\ref{t1}) we obtain 
	\begin{equation}
	\delta u_{\eta \eta \eta \eta \eta}-(\alpha + \delta^3) u_{\eta \eta \eta }- \beta u_{\eta \eta y } - \gamma u_{\eta \eta z}+12\delta (u_{\eta \eta }^2 + u_\eta u_{\eta \eta \eta} )=0. \label{eqn1}
	\end{equation}
	
	Further, by using the logarithmic transformation with the superposition of arbitrary functions as
	\begin{equation}
	u(\eta, y,z) = u_0 (y,z)  +  \left[\log f(\eta , y, z)\right]_\eta, \label{eqn2}
	\end{equation}
	the above equation (\ref{eqn1}) can be deduced into the following bilinear form: 
	\begin{equation}
	(\delta D^4_\eta - \chi D_\eta ^2 - \beta D_\eta D_y - \gamma D_\eta D_z) f \cdot f=0, \label{eqn3}
	\end{equation} 
	where $\chi =(\alpha + \delta^3)$. Here $u_0 (y,z)$ is an arbitrary spatial background, while $f(\eta , y, z)$ is the unknown function to be determined. From the above bilinear form (\ref{eqn3}), one can get different types of nonlinear wave solutions by considering appropriate initial seed solution. Here, we obtain the single solitary wave solutions by taking an exponential test function as the initial seed solution and investigate their dynamics on different backgrounds. Further, the following analysis can be extended to explore multiple soliton dynamics on arbitrary backgrounds in a straightforward manner. Note that Wazwaz studied the multi-soliton solution using trilinear form but without any background \cite{amw4}. However, here, our importance is to focus the effectiveness and consequences of arbitrary spatial backgrounds on solitons.
	
	To construct one soliton solution, we choose the initial seed solution as $f=1+ \varepsilon_1 e^{\Omega_1}$, where $\Omega_1=\theta_1 \eta + \phi_1 y + \psi_1 z+ \delta_1$, and 
	upon its substitution into the bilinear equation \eqref{eqn3} and solving it, we get $\psi_1 = {(\delta \theta_1^3  -\chi \theta_1 - \beta \phi _1)}/{\gamma}$, which results in the following explicit form for $f$:
	\begin{equation}
	f=1+ \varepsilon_1 e^{\theta_1 \eta + \phi_1 y + (\delta \theta_1^3  - \chi \theta_1 - \beta \phi _1)z/{\gamma}+ \delta_1}. \label{eqn5}
	\end{equation}
	Thus the resulting one soliton solution can be written from Eq. (\ref{eqn2}) as 
	\begin{equation} 
	u(\eta, y,z) = u_0 (y,z)  + \frac{\varepsilon_1 \theta_1 e^{\theta_1 \eta + \phi_1 y + (\delta \theta_1^3 - \beta \phi _1 + \chi \theta_1)z/{\gamma}+ \delta_1}}{1+ \varepsilon_1 e^{\theta_1 \eta + \phi_1 y + (\delta \theta_1^3 - \beta \phi _1 + \chi \theta_1)z/{\gamma}+ \delta_1}}. \label{eqn6} 
	\end{equation}
	The above one soliton solution (\ref{eqn6}) contains seven arbitrary constants $\varepsilon_1$, $\theta_1$, $\phi_1$, $\delta_1$, $\alpha$, $\beta$, and $\gamma$ where the first four parameters results exclusively from the solution while the latter three come from the model itself. Apart from these arbitrary constants, we have an additional arbitrary background $u_0 (y,z)$ and it can deliver a rich variety of characteristics across the two spatial dimensions $y$ and $z$. Note that the nature of soliton solution (\ref{eqn6}) we obtained here is nothing but the kink solitons and it carries interesting properties such as amplitude and localization/orientation. 
	\begin{figure}[h]
		\centering\includegraphics[width=0.99\linewidth]{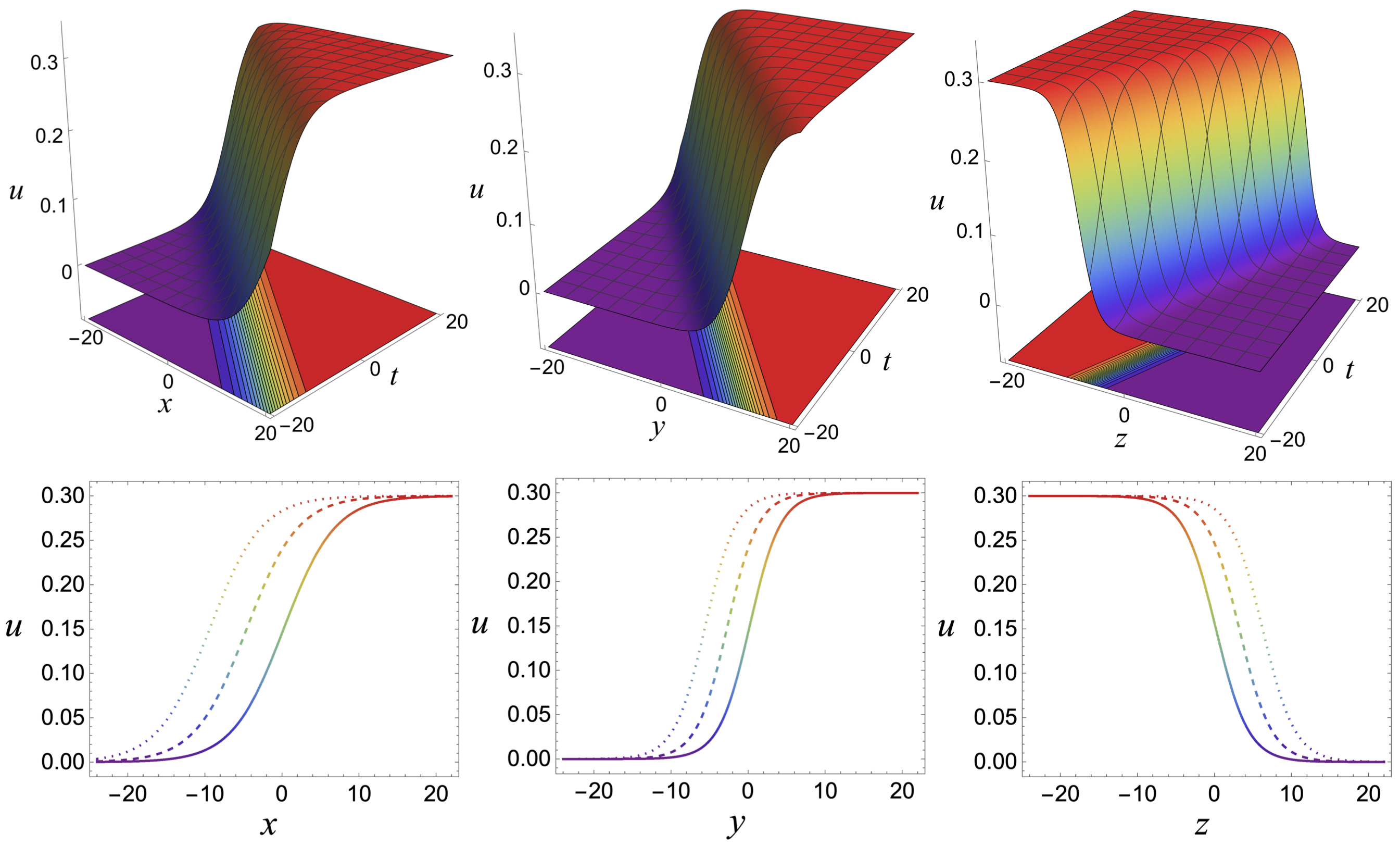} 
		\caption{Evolution of kink-solitons along $x-t$ at $y,z=0.2$ (left panels) and $y-t$ at $x,z=0.2$ (middle panels) and anti-kink-soliton along $z-t$ at $y,x=0.2$ (right panels) obtained through the first-order (one) solution (\ref{eqn6}) for the choice $\delta=1.25$, $\delta_1=1$, $\epsilon_0=0$, $\epsilon_1=1.45$, $\theta_1=0.3$, $\phi_1=0.5$, $\alpha=0.4$, $\beta=0.9$, and $\gamma=1.3$. Bottom panels show kink and anti-kink solitons at different time $t=-1.5$ (solid line), $t=0$ (dashed line), and $t=1.5$ (dotted line).}
		\label{fig-one-sol-1}
	\end{figure} 
	Especially, the solution (\ref{eqn6}) leads to kink soliton (smooth step-like profile) along $x-t$ and $y-t$ while it supports anti-kink soliton (smooth reverse-step-like profile) along $z-t$ due to different spatio-temporal behaviour. 
	From the solution, we can understand that the amplitude of these kink soliton is defined by $\varepsilon_1 \theta_1$, whereas its velocity is characterized by $-\delta$, $-\delta\theta_1/\phi_1$, and $-\delta\theta_1\gamma/(\delta \theta_1^3 - \beta \phi _1 + \chi \theta_1)$ along $x-t$, $y-t$, and $z-t$, respectively. One can tune the nature of resulting kink soliton by using the available arbitrary parameters. For an easy understanding, we have depicted the propagation of such first-order solution taking kink and anti-kink soliton forms in Fig. \ref{fig-one-sol-1}. Also, it shows the projected contour plots and line plots along the spatial dimensions at different times. 
	\begin{figure}[h]
		\centering\includegraphics[width=0.9\linewidth]{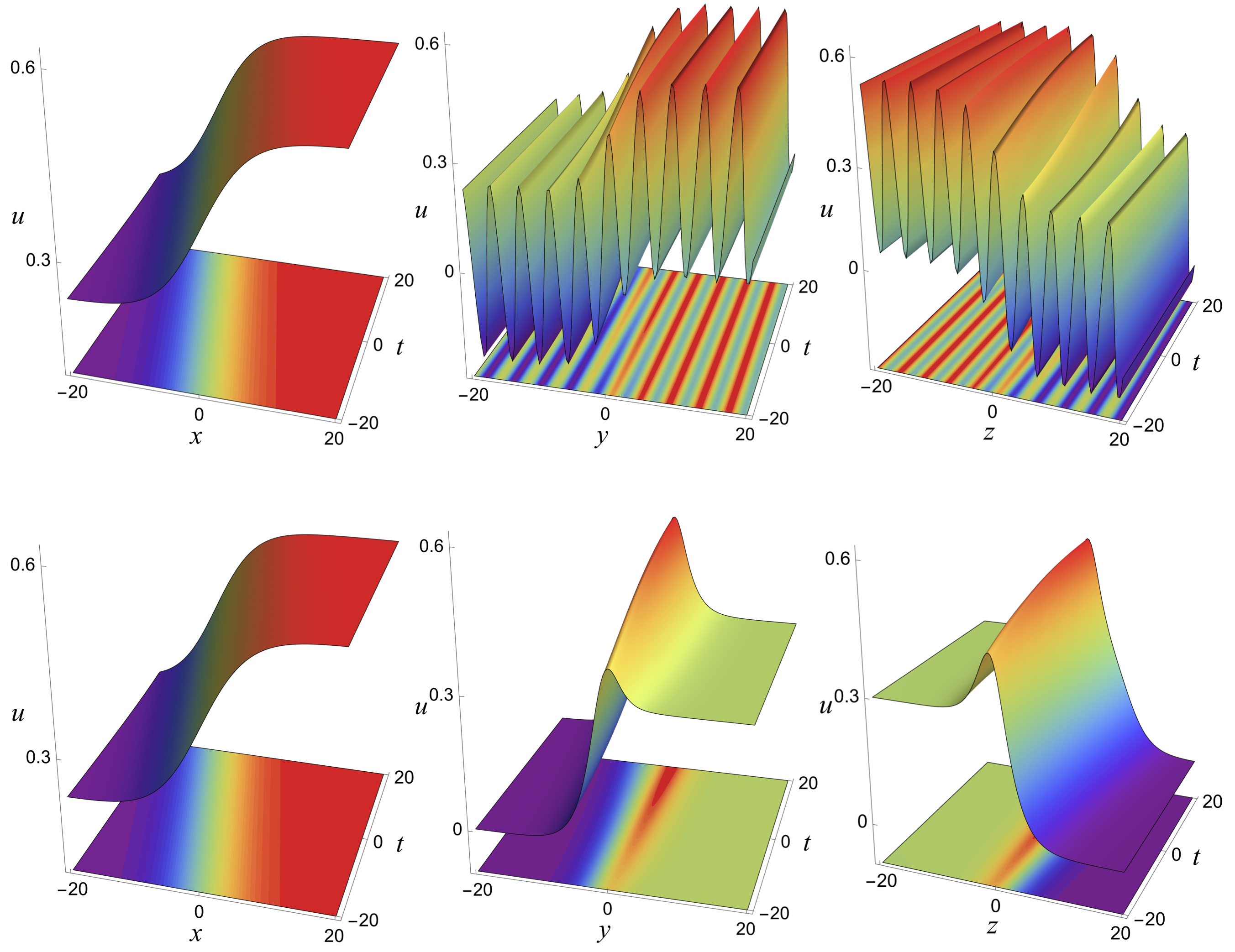}
		\caption{Nature of wide kink-soliton along $x-t$ at $y,z=0.2$ (left panels), narrow kink-soliton along $y-t$ at $x,z=0.2$ (middle panels) and anti-kink-soliton along $z-t$ at $y,x=0.2$ (right panels) obtained through the first-order (one) solution (\ref{eqn6}) for the choice $\delta=1.25$, $\delta_1=1$, $\epsilon_0=0$, $\epsilon_1=1.45$, $\theta_1=0.3$, $\phi_1=0.5$, $\alpha=0.4$, $\beta=0.9$, and $\gamma=1.3$.}
		\label{fig-one-sol-2} 
	\end{figure} 
	
\begin{figure}[h]
		\centering\includegraphics[width=0.91\linewidth]{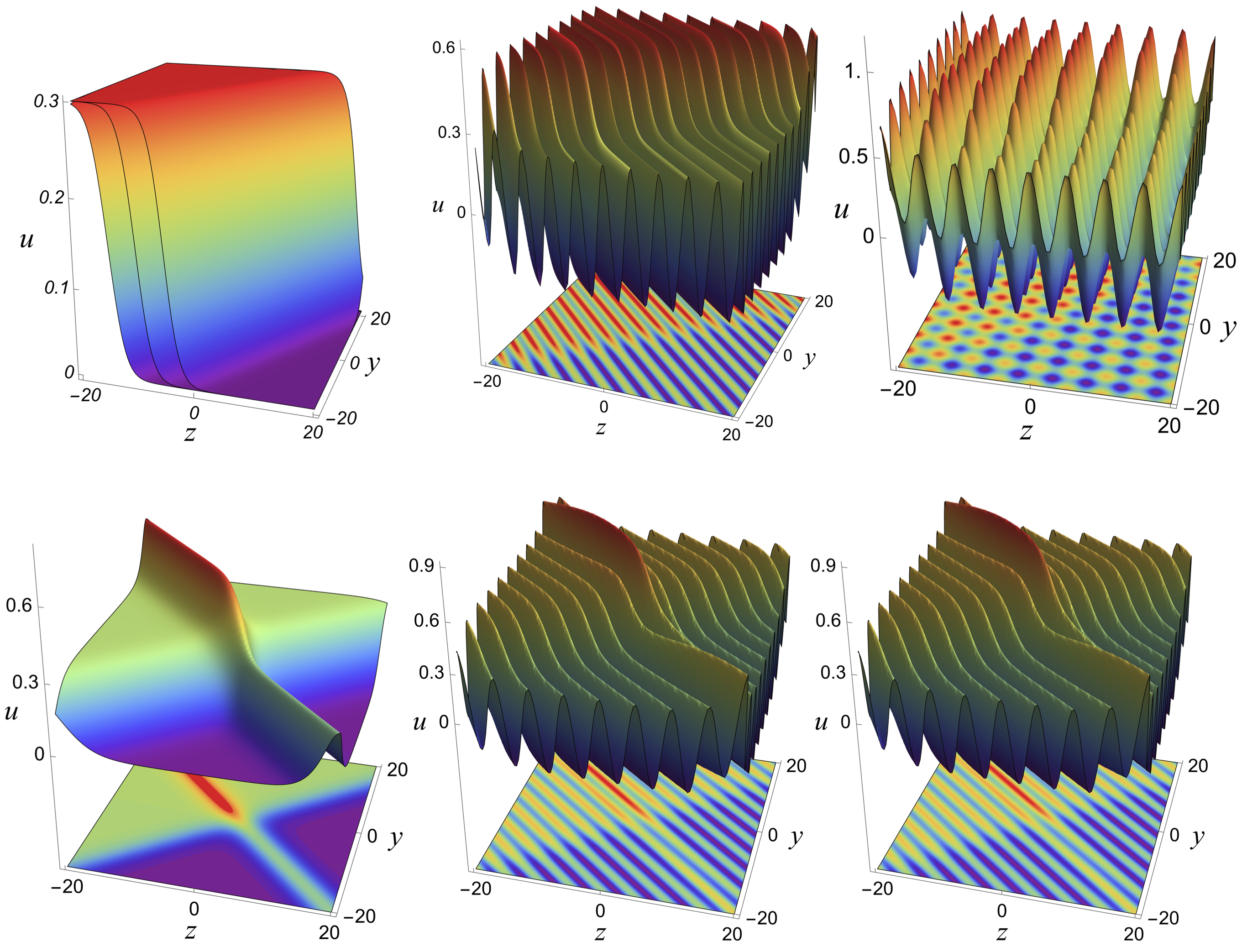}
		\caption{The travelling nature of first-order (one) anti-kink soliton along $y-z$ at different time $t=-8$, $t=0$, and $t=8$ through the solution (\ref{eqn6}) with other parameters as in Fig. \ref{fig-one-sol-1} on zero-background. The influence of (top middle) periodic, (top-right) double-periodic, (bottom-left) bell-type soliton, (bottom-middle)kink-like soliton, and (bottom-right) superposed soliton and periodic type backgrounds on the anti-kink soliton along $y-z$ at $t=0$.}
		\label{fig-one-sol-3}
	\end{figure}

	As discussed in the case of lumps, here in the kink solitons also the  arbitrary spatially-varying backgrounds introduces different manifestation of the solitons. Particularly, the periodic background arising for the choice  either $p_1\neq 0$, $p_2=0$, $q_1=0$ or $p_1=0$, $p_2\neq 0$, $q_2=0$ influences the kink soliton to transform into a periodically oscillating kink and anti-kink waves as demonstrated in the top panels of Fig. \ref{fig-one-sol-2} in both $y-t$ and $z-t$ planes. On the other hand, the  localized $sech$ type background ($p_1=0$, $p_2\neq 0$, $q_2=1$) gives rise to the coexistence of bell soliton on top of the narrowing or steepening kink and anti-kink solitons, which we have shown in the bottom panels of Fig. \ref{fig-one-sol-2}. Also, based on the localization of the kink and bell solitons, this steepening effect gets altered. On contrary to the modulations in $y-t$ and $z-t$ planes, the kink solitons not undergo any significant alteration in their dynamics along $x-t$ plane due to the  nature of background function $u_0(y,z)$. 
	Considering this factor, proceeding further, we can observe different other features in the evolution of kink solitons along the $y-z$ plane which are depicted in Fig. \ref{fig-one-sol-3}. Especially, the anti-kink soliton shown in the top-left panel transforms into periodically oscillating kink solitons when $p_1\neq 0$, $p_2=0$, $q_1=0$ or $p_1=0$, $p_2\neq 0$, $q_2=0$ (top-middle panel), double-periodic kink soliton for $p_1,p_2\neq 0$, $q_1,q_2=0$ (top-right panel), and an elastic interaction of kink and bell type solitons when the background is chosen as $p_1=0$, $p_2\neq 0$, $q_2=1$ (bottom-left panel). Additionally, when the background function is considered as a combined periodic and bell waves, one shall identify their superposed effect on the kink soliton as given in the bottom-middle and bottom-right panel of Fig. \ref{fig-one-sol-3}.

	\section{Bilinear B\"acklund Transformation}\label{sec5}
	In this section, we obtained the bilinear B\"acklund transformation of the considered model using the exchange identities of the Hirota's $D$ operator~\cite{Hirota-book,ysh,yxm}. By adapting this alternate approach different types of explicit solutions can be constructed. For this purpose, we introduce the following equation for the functions $f$ and $g$ appearing in the bilinear equation (\ref{eqn3}):
	\bea
	&&P=[  ( \delta D_\eta ^4 - \chi D_\eta ^2 - \beta D_{\eta}D_y - \gamma D_\eta D_Z  ) g \cdot g  ] f^2 - g^2  [  (  \delta D_{\eta}^4 - \chi D_{\eta }^2 - \beta D_{\eta }D_y-\gamma D_\eta D_Z ) f \cdot f  ] =0,\qquad  \label{eq1}\\
	&&\quad =\delta [ (D_{\eta}^4 g \cdot g ) f^2 - g^2(D_\eta ^4 f \cdot f ) ] - \chi [ (D_{\eta}^2 g \cdot g ) f^2 - g^2(D_\eta ^2 f \cdot f ) ] \nonumber\\ 
	&& \qquad - \beta [ (D_{\eta} D_y  g \cdot g ) f^2 - g^2(D_\eta D_y f \cdot f ) ] - \gamma [ (D_{\eta} D_z g \cdot g ) f^2 - g^2(D_\eta D_y f \cdot f ) ]=0,  \label{eq2}   
	\eea 
	where $g$ is real differential function of $\eta, y$ and $z$. 
	Using the exchange identities for the Hirota $D$ operator \cite{Hirota-book},
	\bes \begin{align}
	(D_{\eta}^2 g \cdot g ) f^2 - g^2 (D_\eta  ^2 f \cdot f ) & = 2 D_\eta (D_\eta g \cdot f ) \cdot (fg), \label{eq3a}\\
	(D_\eta D_y g \cdot g ) f^2 - g^2 (D_\eta D_y f \cdot f ) & = 2 D_\eta (D_y g \cdot f ) \cdot (fg), \label{eq3b}\\
	(D_{\eta}^4 g \cdot g ) f^2 - g^2 (D_\eta  ^4 f \cdot f ) & = 2 D_\eta (D_\eta ^3  g \cdot f ) \cdot (fg) - 6 D_\eta (D_\eta ^2 g \cdot f ) \cdot (D_\eta g \cdot f). \label{eq3c}
	\end{align} \label{bilid} \ees 
	By applying the above bilinear identities (\ref{bilid}) in Eq. \eqref{eq2}, we can arrive at the following form:
	\bea 
	&&P= \delta [ 2 D_\eta (D_\eta ^3 g \cdot f ) \cdot (fg) - 6 D_\eta (D_\eta ^2 g \cdot f ) \cdot (D_\eta g\cdot f)] - \chi [2 D_\eta (D_\eta g \cdot f )\cdot (fg) ]\nonumber\\ 
	&&\qquad  - \beta [2 D_\eta (D_y g \cdot f) \cdot (fg)] - \gamma [2 D_\eta (D_z g\cdot f ) \cdot (fg)] =0. \label{eq4}
	\eea 
	Without loss of generality, assuming that $D_\eta ^2 g \cdot f = \varepsilon _1 fg $, we obtain 
	\begin{align}
	2 D_\eta [(\delta D_\eta ^3 + 3 \varepsilon_1 \delta D_\eta - \chi D_\eta - \beta D_y - \gamma D_z) g \cdot f ](fg) =0. \label{eq5}
	\end{align}
	Finally, we have the following bilinear B\"acklund transformation:
	\begin{subequations} 
		\bea
		&&u= u_0 (y,z) +  (\log f)_\eta, \label{eq6a}\\
		&&(\delta D_\eta ^3 + 3 \varepsilon_1 \delta D_\eta  -\chi D_\eta - \beta D_y - \gamma D_Z) g \cdot f =0, \label{eq6b}\\
		&&(D_\eta ^2 - \varepsilon_1 ) g \cdot f =0. \label{eq6c}
		\eea
	\end{subequations}
	By adopting the above  bilinear B\"acklund transformation, we can construct different classes of nonlinear wave solutions on arbitrary spatial background with suitably chosen form of initial functions for $g$ and $f$. For completeness and demonstration purpose, we have constructed hyperbolic and periodic wave solutions in the forthcoming part. 
	\subsection{Hyperbolic Wave Solution }
	We choose the seed functions $f=1$ and $g = \cosh (m_1 \eta +\eta _1 y + p_1 z + r_1 )$ and solving \eqref{eq6b}-\eqref{eq6c}, we get 
	$p_1 = ({4m_1 ^3 \delta - m_1 \chi - \eta_1 \beta  })/{\gamma}.$  
	This gives us the explicit form of $g$ and $f$ as given below.
	\begin{align}
	f=1, \quad g =\cosh \left(m_1 \eta + n_1 y +  {(4m_1 ^3 \delta - m_1 \chi - \eta_1 \beta  )z}/{\gamma}+r_1\right). \label{eq8}
	\end{align}
	Thus, the final hyperbolic solution of the considered model Eq. (\ref{t1}) is obtained as 
	\begin{align}
	u(x,y,z,t) & =u_0 (y,z)  + m_1 \tanh \left( (x+\delta t)m_1+n_1 y + {(4m_1 ^3 \delta - m_1 (\alpha+\delta^3) - \eta_1 \beta  )z}/{\gamma}+r_1 \right). \label{eq9}
	\end{align}
	We can obtain another hyperbolic wave solution, by choosing 
	$f=1$ and $g (\eta , y , z)=\sinh (m_1 \eta + \eta _1 y + p_1 z +r_1)$, 
	and solving \eqref{eq6b}-\eqref{eq6c} which would give the same { $p_1$ as above}. Then, the final form of the hyperbolic wave solution of the considered model Eq. (\ref{t1}) is obtained as 
	\begin{align}
	u(x,y,z,t) & =u_0 (y,z)  + m_1 \coth \left( (x+\delta t)m_1+n_1 y + {(4m_1 ^3 \delta - m_1 (\alpha+\delta^3) - \eta_1 \beta  )z}/{\gamma}+r_1 \right). \label{eq91}
	\end{align}
	
	\subsection{Periodic Wave Solution}
	Here, we have obtained periodic wave solutions by choosing $f=1$, $g = \cos (m_1 \eta + \eta _1 y + p_1 z +r_1) $,  and solving \eqref{eq6b}-\eqref{eq6c}, we get 
	$p_1 = - ({(4m_1^3 \delta+m_1 \chi + \eta _1 \beta)})/{\gamma}$.
	Finally, the resultant periodic wave solution  of the considered soliton model (\ref{t1}) is obtained as
	\begin{align}
	u(x,y,z,t) & =u_0 (y,z)  - m_1 \tan \left( (x+\delta t)m_1+n_1 y - {(4m_1 ^3 \delta + m_1 (\alpha+\delta^3) + \eta_1 \beta  )z}/{\gamma}+r_1 \right). \label{eq92}
	\end{align}
	
	Similarly, we can have alternate form of periodic wave solution of the model (\ref{t1}), by choosing  $f=1, g = \sin (m_1 \eta + \eta _1 y + p_1 z +r_1) $, and it can be written as follows.
	\begin{align}
	u(x,y,z,t) & =u_0 (y,z)  + m_1 \cot \left( (x+\delta t)m_1+n_1 y - {(4m_1 ^3 \delta + m_1 (\alpha+\delta^3) + \eta_1 \beta  )z}/{\gamma}+r_1 \right). \label{eq93}
	\end{align}
	It is necessary to highlight the fact that by using different initial seed solutions, the above  bilinear B\"acklund transformation can be extended straightforwardly to obtain different other wave solutions. Considering the length of the article, we refrain from presenting such detailed class of solutions and further discussion here. Also, note that the above periodic and hyperbolic solutions exhibit arbitrary spatial background $u_0 (y,z) $ which again give us fruitful dynamics in the underlying wave structures. 
	
	\section{Conclusions}\label{sec-conclusion}
	{  To summarize the present work, we have investigated an integrable (3 $+$ 1)-dimensional nonlinear equation~\eqref{t1} describing the evolution of {oceanic waves} with higher-order temporal dispersion for the dynamics of the lump and solitons on different spatial backgrounds. Particularly, we have constructed explicit solutions for lump and kink-soliton of the considered model through a systematic analysis using its trilinear and bilinear formulations, respectively, along with suitable forms of polynomial and exponential type initial seed solutions. Our analysis shows that the lump solution results into a well-localized profile with a simultaneous existence of spiking/peak and declining/dip (coupled bright-dark peaks) structure on a constant background. However, the obtained soliton solution leads to a step-like kink and anti-kink patterns along different spatio-temporal domain. 
		When the arbitrary spatial background function arising in the solution is
		incorporated (non-zero), the lump wave results into a number of interesting
		wave phenomena. Especially, the periodic and localized type arbitrary spatial
		backgrounds generate (i) coexistence of the lump on periodic wave, (ii) breather
		formation on periodic background wave, (iii) spiking on top of bell \& kink
		solitons, (iv) interaction of line soliton-like nature, and (v) interacting
		lump with kink wave. Additionally, the incorporation of these periodic and
		localized soliton spatial backgrounds play a crucial role in the transition of
		kink solitons into manipulated interaction waves with bell type soliton,
		periodic waves, double-periodic waves with substantial changes in their
		dynamics. For a better understanding of the resulting dynamics, we have
		provided a categorical discussion and clear graphical demonstration for lump
		and solitons on both constant and spatially-varying backgrounds. Further, we
		have obtained periodic and hyperbolic solutions with arbitrary spatial
		backgrounds for the considered model \eqref{t1} through bilinear B\"acklund
		transformation. The obtained results will be an important addition along the
		context of nonlinear wave manipulation in higher-dimensional models due to
		controllable backgrounds. The present investigation shall also be extended to
		several other single and multi-component real and complex soliton models
		towards enhanced understanding of the dynamical characteristics of exotic
		nonlinear waves like rogue waves and dromions on variable
		background~\cite{ma1,ma2}. }  \\ 
	
	\setstretch{1.0}	
	\noindent{\bf Acknowledgments}\\
	One of the authors SS acknowledges Ministry of Education (MoE), India, for the financial support through institute (National Institute of Technology, Tiruchirappalli) fellowship. The research work of K Sakkaravarthi is supported by the Young Scientist Training (YST) program of the Asia-Pacific Center for Theoretical Physics (APCTP), Pohang-si, Gyeongsangbuk-do, South Korea. The APCTP is supported by the Korean Government through the Science and Technology Promotion Fund and Lottery Fund.
	{ The authors thank the handling editor and anonymous reviewers for providing valuable comments and suggestions that helped immensely to improve the quality of the manuscript.}\\
	
	\noindent{\bf CRediT Author Contribution Statement}\\ {\bf Sudhir Singh}: Conceptualization, Methodology, Writing - Original Draft Preparation, Writing - Review \& Editing. {\bf K. Sakkaravarthi}: Formal Analysis, Investigation, Visualization, Writing - Original Draft Preparation, Writing - Review \& Editing. {\bf K. Murugesan}: Resources, Writing - Review \& Editing, Funding acquisition, Supervision. \\
	
	\noindent{\bf Declaration of Competing Interest}\\ The authors declare that they have no known competing financial interests or personal relationships that could have appeared to influence the work reported in this paper.
	
	\setstretch{0.950}
	

\begin{thebibliography}{9}
		\bibitem{Yang-book}	
		J. Yang, Nonlinear Waves in Integrable and Nonintegrable Systems (SIAM, Philadelphia, 2010).
		
		\bibitem{ML-book}	
		M. Lakshmanan, S. Rajasekar, Nonlinear Dynamics: Integrability, Chaos and Patterns (Springer, Berlin, 2003).
		
		\bibitem{bgo}
		B. Guo, X-F. Pang,  Y-F. Wang, N. Liu, Solitons (De Gruyter, Berlin, 2018).
		
		\bibitem{bgu}
		B. Guo, L. Tian, Z. Yan, L. Ling, Y-F. Wang, Rogue waves: Mathematical theory and applications
		in physics (De Gruyter, Berlin, 2017).
		
		\bibitem{ksa}
		K. Sakkaravarthi, T. Kanna, R. Babu Mareeswaran, Higher-order Optical Rogue Waves in Spatially Inhomogeneous Multimode Fiber, Phys. D: Nonlinear Phenom. 435 (2022) 133285. 
		
		\bibitem{boris}
		E. Kengne, W.M. Liu, B.A. Malomed, Spatiotemporal engineering of matter-wave solitons in Bose-Einstein condensates, Phys. Rep. 899 (2021) 1–62. 
		
		{ 
			\bibitem{akm-rogue}
			N. Akhmediev, A. Ankiewicz, M. Taki, Waves that appear from nowhere and disappear without a trace, Phys. Lett. A 373 (2009) 675–678.
		}
		
		\bibitem{zba}
		Z. Bai, G. Huang, Plasmon dromions in a metamaterial via plasmon-induced transparency, Phys. Rev. A 93
		(2016) 013818.
		
		\bibitem{mra}
		M. R. Alam, Dromions of flexural-gravity waves, J. Fluid. Mech. 719 (2013) 1-13.
		
		\bibitem{xwj}
		X-W. Jin, S-J. Shen,  Z-Y. Yang, J. Lin, Magnetic lump motion in saturated ferromagnetic films, Phys. Rev. E 105 (2022) 014205.
		
		\bibitem{wx1}
		W-X. Ma, Lump solutions to the Kadomtsev-Petviashvili equation, Phys. Lett. A 379 (2015) 1975-1978.
		
		\bibitem{djf}
		D. J. Frantzeskakis, T. P. Horikis, A. S. Rodrigues, P. G. Kevrekidis, R. Carretero-Gonz\'alez, and J. Cuevas-Maraver, Hydrodynamics and two-dimensional dark lump solitons for polariton superfluids, Phys. Rev. E 98 (2018) 022205.
		
		\bibitem{fba}
		F. Baronio, S. Wabnitz,  Y. Kodama, Optical Kerr Spatiotemporal Dark-Lump Dynamics of Hydrodynamic Origin, Phys. Rev. Lett. 116 (2016) 173901.
		
		\bibitem{hdg}
		H-D. Guo, T-C. Xia,  B-B. Hu, Dynamics of abundant solutions to the (3+1)-dimensional generalized Yu–Toda–Sasa–Fukuyama equation, Appl. Math. Lett. 105 (2020) 106301. 
		
		\bibitem{gqx}
		G-Q Xu, A-M. Wazwaz, Characteristics of integrability, bidirectional solitons and localized solutions for a (3 + 1)-dimensional generalized breaking soliton equation, Nonlinear Dyn. 96 (2019) 1989–2000. 
		
		\bibitem{ylm}
		Y-L. Ma,  A.M. Wazwaz, B-Q. Li, New extended Kadomtsev-Petviashvili equation: multiple soliton solutions, breather, lump and interaction solutions, Nonlinear Dyn. 104 (2021) 1581–1594.
		
		\bibitem{lka}
		L. Kaur,  A.M. Wazwaz, Lump, breather and solitary wave solutions to new reduced form of the generalized BKP equation, Int. J. Numer. Meth. Heat Fluid Flow 29 (2019) 569-579.
		
		\bibitem{amw1}
		A.M. Wazwaz, A new fifth-order nonlinear integrable equation: multiple soliton solutions, Phys. Scr. 83 (2011) 015012.
		
		\bibitem{amw2}
		A.M. Wazwaz, A new generalized fifth-order nonlinear integrable equation. Phys. Scr. 83 (2011) 035003.
		
		
		\bibitem{gwa}
		G. Wang, X.  Liu, Y. Zhang, Symmetry reduction, exact solutions and conservation laws of a new fifth-order nonlinear integrable equation, Commun. Nonlinear Sci. Numer. Simul. 18 (2013) 2313-2320.
		
		
		\bibitem{amw3}
		A. M. Wazwaz,  Kink solutions for three new fifth order nonlinear equations, Appl. Math. Model. 38 (2014) 110. 
		
		\bibitem{gqx1}
		G-Q. Xu, A. M.  Wazwaz, Bidirectional solitons and interaction solutions for a new integrable fifth-order nonlinear equation with temporal and spatial dispersion, Nonlinear Dyn. 101 (2020) 581-595. 
		
		\bibitem{rip1}
		M. Alquran, Physical properties for bidirectional wave solutions to a generalized fifth-order equation with third-order time-dispersion term, Res. Phys. 28 (2021) 104577.
		
		\bibitem{jgr}
		J. Gratus, P. Kinsler, M. W.  McCall,  On spacetime transformation optics: temporal and spatial dispersion. New J. Phys. 18 (2016) 123010. 
		
		\bibitem{csc}
		C. Schelte, A.  Pimenov, A. G.  Vladimirov, Tunable Kerr frequency combs and temporal localized states in time-delayed Gires-Tournois interferometers, Optics Lett. 44 (2019) 4925–4928.
		
		\bibitem{amw4}
		A-M. Wazwaz, New (3+1)-dimensional Painlev\'e integrable fifth-order equation with third-order temporal dispersion, Nonlinear Dyn. 106 (2021) 891-897.
		
		{ 
			\bibitem{cp1}
			X-T. Gao, B. Tian, Water-wave studies on a (2+1)-dimensional generalized variable-coefficient Boiti-Leon-Pempinelli system, Appl. Math. Lett. 128 (2022) 107858. 
			
			\bibitem{cp2}
			X-Y. Gao, Y-J. Guo, W-R. Shan, Taking into consideration an extended coupled (2+1)-dimensional Burgers system in oceanography, acoustics and hydrodynamics, Chaos Solitons Fractals 161 (2022) 112293.
			
			\bibitem{cp3} 
			X-Y. Gao, Y-J. Guo, W-R. Shan, Similarity reductions for a generalized (3+1)-dimensional variable-coefficient B-type Kadomtsev-Petviashvili equation in fluid dynamics, Chin. J. Phys. 77 (2022) 2707-2712.
			
			\bibitem{cp4} 
			X-Y. Gao, Y-J Guo, W-R. Shan, Optical waves/modes in a multicomponent inhomogeneous optical fiber via a three-coupled variable-coefficient nonlinear Schr\"odinger system, Appl. Math. Lett. 120 (2021) 107161. }
		
		\bibitem{amwb}
		A-M. Wazwaz, Partial Differential Equations and Solitary Waves Theory  (Springer, Heidelberg, 2009).
		
		\bibitem{Hirota-book}
		R. Hirota, The direct method in soliton theory (Cambridge University Press, Cambridge, 2004).
		
		{\cb	\bibitem{abc4}
			Z. Zhang, X. Yang, B. Li, A-M. Wazwaz, Q. Guo, Generation mechanism of high-order rogue waves via the improved long-wave limit method: NLS case, Phys. Lett. A. 450 (2022) 128395.}
		
		{ 
			\bibitem{cl1}
			X-Y. Gao, Y-J. Guo, W-R. Shan, Bilinear forms through the binary Bell polynomials, $N$ solitons and B\"acklund transformations of the Boussinesq-Burgers system for the shallow water waves in a lake or near an ocean beach, Commun. Theor. Phys. 72 (2020) 095002. 
			
			\bibitem{cl2}
			X-Y. Gao, Y-J. Guo, W-R. Shan, Regarding the shallow water in an ocean via a Whitham-Broer-Kaup-like system: hetero-B\"acklund transformations, bilinear forms and $M$ solitons, Chaos Solitons Fractals 162 (2022) 112486.
			
			\bibitem{cl3} T-Y. Zhou, B. Tian, Y-Q. Chen, Y. Shen, Painlev\'e analysis, auto-B\"acklund transformation and analytic solutions of a (2+1)-dimensional generalized Burgers system with the variable coefficients in a fluid, Nonlinear Dyn. 108 (2022) 2417–2428. 
			
			\bibitem{cl4} 
			X-Y. Gao, Y-J. Guo, W-R. Shan, T-Y. Zhou, M. Weng, D-U. Yang, In the atmosphere and oceanic fluids: scaling transformations, bilinear forms, B\"acklund transformations and solitons for a generalized variable-coefficient Korteweg-de Vries-modified Korteweg-de Vries equation, China Ocean Eng. 35 (2021) 518-530.
			
		}
		{\cb	
			\bibitem{abc1}
			Z. Zhang, B. Li, J. Chen, Q. Guo, Y. Stepanyants, Peculiarities of resonant interactions of lump chains within the KP1 equation, Phys. Scr. 97 (2022) 115205.}
		
		{\cb		\bibitem{abc2}
			Z. Zhang, B. Li, J. Chen, Q. Guo, Y. Stepanyants, Degenerate lump interactions within the Kadomtsev-Petviashvili equation, Commun. Nonlinear Sci. Numer. Simul. 112  (2022) 106555.} 
		
		{\cb	\bibitem{abc3}
			Z. Zhang, X. Yang, B. Li,
			Q. Guo, Y. Stepanyants, Multi-lump formations from lump chains and plane solitons in the KP1 equation, Nonlinear Dyn (2022) https://doi.org/10.1007/s11071-022-07903-8.}	
		
		{ 		
			\bibitem{ch1}
			Z. Zhang, C. Luo, Z. Zhao, Application of probabilistic method in maximum tsunami height prediction considering stochastic seabed topography, Nat. Hazards 104 (2020) 2511-2530.
			
			\bibitem{ch2}
			G. Dematteis, T. Grafke, E. Vanden-Eijnden, Rogue waves and large deviations in deep sea, PNAS 115 (2018) 855-860.
			
			\bibitem{ch3}
			Y. Wang, H. Wang, B. Zhou, H. Fu, Multi-dimensional prediction method based on Bi-LSTMC for ship roll, Ocean Eng. 242 (2021) 110106. 
			
			\bibitem{ch4}
			Z. Li, M-A. Meier, E. Hauksson, Z. Zhan, J. Andrews, Machine Learning Seismic Wave Discrimination: Application to Earthquake Early Warning, Geophys. Res. Lett. 45 (2018) 4773-4779.
			
			\bibitem{ch5}
			S.C. James, Y. Zhang, F. O\'Donncha, A machine learning framework to forecast wave conditions, Coast. Eng. 137 (2018) 1–10.} 
		
		
		
		
		
		
		
		
		
		\bibitem{ksjpa20}
		R.B. Mareeswaran, K. Sakkaravarthi, T. Kanna, Manipulation of vector solitons in a system of inhomogeneous coherently coupled nonlinear Schr\"odinger
		models with variable nonlinearities, J. Phys. A: Math. Theor. 53 (2020) 415701. 
		
		\bibitem{ksps20}
		K. Sakkaravarthi, R. B. Mareeswaran, T. Kanna, Engineering optical rogue waves and breathers in a coupled nonlinear Schr\"odinger system with four-wave mixing effect, Physica Scr. 95 (2020) 095202. 
		
		{ 
			\bibitem{tkan}
			T. Kanna, R. B. Mareeswaran, F. Tsitoura, H. E. Nistazakis, D. J. Frantzeskakis, Non-autonomous bright–dark solitons and Rabi oscillations in multi-component Bose–Einstein condensates, J. Phys. A: Math. Theor. 46 (2013) 475201.
			
			\bibitem{mani1}
			K. Manikandan, N. Vishnu Priya, M. Senthilvelan, R. Sankaranarayanan. Higher-order matter rogue waves in two-component Bose-Einstein condensates, Waves Random Complex Media 32 (2022) 867–886.
			
			\bibitem{mani2}
			N. Sinthuja, K. Manikandan, M. Senthilvelan, Rogue waves on the double-periodic background in Hirota equation, Eur. Phys. J. Plus 136 (2021) 305.
		}
		
		\bibitem{tk1}
		T. Kanna, R.B. Mareeswaran, K. Sakkaravarthi, Non-autonomous bright matter wave solitons in spinor Bose–Einstein condensates, Phys. Lett. A 378 (2014) 158-170. 
		
		\bibitem{tk2}
		T. Kanna, A. Annamalar Sheela, R. Babu Mareeswaran, Spatially modulated two-and three-component Rabi-coupled Gross–Pitaevskii systems, J. Phys. A: Math. Theor. 52 (2019) 375201. 
		
		\bibitem{pfh}
		J-G. Liu, W-H. Zhu, Breather wave solutions for the generalized shallow water wave equation with variable coefficients in the atmosphere, rivers, lakes and oceans, Comput. Math. Appl. 78 (2019) 848-856.
		
		{\cb \bibitem{epjp19}
			J-G Liu, W-H Zhu, Y. He, Z-Q Lei, Characteristics of lump solutions to a (3+1)-dimensional variable-coefficient generalized shallow water wave equation in oceanography and atmospheric science, Eur. Phys. J. Plus 134 (2019) 385. }
		
		\bibitem{ssi}
		S. Singh, K. Sakkaravarthi, K. Murugesan, Localized nonlinear waves on spatio-temporally controllable backgrounds for a (3+1)-dimensional Kadomtsev-Petviashvili-Boussinesq model in water waves, Chaos Solitons Fractals 155 (2022) 111652.
		
		\bibitem{jgl}
		P-F. Han, Y. Zhang, Linear superposition formula of solutions for the extended (3+1)-dimensional shallow water wave equation, Nonlinear Dyn. 109 (2022) 1019–1032.
		
		\bibitem{pfha}
		P-F. Han, T. Bao, Interaction of multiple superposition solutions for the (4+1)-dimensional Boiti-Leon-Manna-Pempinelli equation, Nonlinear Dyn.  105 (2021) 717–734.
		
		{\cb \bibitem{epjp21}
			P-F Han, T. Bao, Dynamic analysis of hybrid solutions for the new (3+1)-dimensional Boiti–Leon–Manna–Pempinelli equation with time-dependent coefficients in incompressible fluid, Eur. Phys. J. Plus 136 (2021) 925.}
		
		\bibitem{ysh}
		Y. Shen, B. Tian, Bilinear auto-B\"acklund transformations and soliton solutions of a
		(3+1)-dimensional generalized nonlinear evolution equation for the shallow water waves, Appl. Math. Lett. 122 (2021)  107301.
		
		\bibitem{yxm}
		Y-X. Ma, B. Tian, Q-X. Qu,  C-C Wei, X. Zhao, B\"acklund transformations, kink soliton, breather- and travelling-wave solutions for a (3+1)-dimensional B-type
		Kadomtsev-Petviashvili equation in fluid dynamics, Chinese J. Phys. 73 (2021) 600-612.
		
		
		{ 		\bibitem{ma1}
			C-J. Cui, X-Y. Tang, Y-J. Cui, New variable separation solutions and wave interactions for the (3+1)-dimensional Boiti-Leon-Manna-Pempinelli equation, Appl. Math. Lett. 102 (2020) 106109. 
			
			\bibitem{ma2}  
			L. Huang, On the dynamics of localized excitation wave solutions to an
			extended (3+1)-dimensional Jimbo-Miwa equation, Appl. Math. Lett. 121 (2021) 107501. 
			
		}
		
		
		
		
	\end{thebibliography}
\end{document}